\documentclass{ws-procs9x6-cpt25}
\begin{document}
	
	\newcommand{\refeq}[1]{(\ref{#1})}
	\def\etal {{\it et al.}}
	
	\def\al{\alpha}
	\def\be{\beta}
	\def\ga{\gamma}
	\def\de{\delta}
	\def\ep{\epsilon}
	\def\ve{\varepsilon}
	\def\ze{\zeta}
	\def\et{\eta}
	\def\th{\theta}
	\def\vt{\vartheta}
	\def\io{\iota}
	\def\ka{\kappa}
	\def\la{\lambda}
	\def\vpi{\varpi}
	\def\rh{\rho}
	\def\vr{\varrho}
	\def\si{\sigma}
	\def\vs{\varsigma}
	\def\ta{\tau}
	\def\up{\upsilon}
	\def\ph{\phi}
	\def\vp{\varphi}
	\def\ch{\chi}
	\def\ps{\psi}
	\def\om{\omega}
	\def\Ga{\Gamma}
	\def\De{\Delta}
	\def\Th{\Theta}
	\def\La{\Lambda}
	\def\Si{\Sigma}
	\def\Up{\Upsilon}
	\def\Ph{\Phi}
	\def\Ps{\Psi}
	\def\Om{\Omega}
	\def\cA{{\cal A}}
	\def\cB{{\cal B}}
	\def\cC{{\cal C}}
	\def\cE{{\cal E}}
	\def\cl{{\mathcal L}}
	\def\cL{{\mathcal L}}
	\def\cO{{\cal O}}
	\def\cP{{\cal P}}
	\def\cR{{\cal R}}
	\def\cV{{\cal V}}
	\def\mn{{\mu\nu}}
	
	\def\fr#1#2{{{#1} \over {#2}}}
	\def\half{{\textstyle{1\over 2}}}
	\def\quar{{\textstyle{1\over 4}}}
	\def\frac#1#2{{\textstyle{{#1}\over {#2}}}}
	
	\def\vev#1{\langle {#1}\rangle}
	\def\bra#1{\langle{#1}|}
	\def\ket#1{|{#1}\rangle}
	\def\bracket#1#2{\langle{#1}|{#2}\rangle}
	\def\expect#1{\langle{#1}\rangle}
	\def\norm#1{\left\|{#1}\right\|}
	\def\abs#1{\left|{#1}\right|}
	
	\def\lsim{\mathrel{\rlap{\lower4pt\hbox{\hskip1pt$\sim$}}
			\raise1pt\hbox{$<$}}}
	\def\gsim{\mathrel{\rlap{\lower4pt\hbox{\hskip1pt$\sim$}}
			\raise1pt\hbox{$>$}}}
	\def\sqr#1#2{{\vcenter{\vbox{\hrule height.#2pt
					\hbox{\vrule width.#2pt height#1pt \kern#1pt
						\vrule width.#2pt}
					\hrule height.#2pt}}}}
	\def\square{\mathchoice\sqr66\sqr66\sqr{2.1}3\sqr{1.5}3}
	
	\def\prt{\partial}
	
	\def\pt#1{\phantom{#1}}
	\def\ni{\noindent}
	\def\ol#1{\overline{#1}}
	
	\newcommand{\ttens}{\ensuremath{t^{\kappa\lambda\mu\nu}}}
	\newcommand{\uu}{\ensuremath{u}}
	
	\def\sss{s^{\mu\nu}}
	\def\ttt{t^{\ka\la\mu\nu}}
	
	\def\sb{\overline{s}}
	\def\tb{\overline{t}}
	\def\ub{\overline{u}}
	
	\def\stw{\tilde{s}}
	\def\ttw{\tilde{t}}
	\def\utw{\tilde{u}}
	\def\Btw{\tilde{B}}
	
	\def\hsy{h_{\mu\nu}}  
	\def\nsy{\et_{\mu\nu}}

	\def\nsc#1#2#3{\om_{#1}^{{\pt{#1}}#2#3}}
	\def\lsc#1#2#3{\om_{#1#2#3}}
	\def\usc#1#2#3{\om^{#1#2#3}}
	\def\lulsc#1#2#3{\om_{#1\pt{#2}#3}^{{\pt{#1}}#2}}
	
	\def\tor#1#2#3{T^{#1}_{{\pt{#1}}#2#3}}
	
	\def\vb#1#2{e_{#1}^{{\pt{#1}}#2}}
	\def\ivb#1#2{e^{#1}_{{\pt{#1}}#2}}
	\def\uvb#1#2{e^{#1#2}}
	\def\lvb#1#2{e_{#1#2}}
	
	\def\barvb#1#2{\bar e_{#1}^{{\pt{#1}}#2}}
	\def\barivb#1#2{\bar e^{#1}_{{\pt{#1}}#2}}
	\def\baruvb#1#2{\bar e^{#1#2}}
	\def\barlvb#1#2{\bar e_{#1#2}}
	
	\newcommand{\beq}{\begin{equation}}
		\newcommand{\eeq}{\end{equation}}
	\newcommand{\bea}{\begin{eqnarray}}
		\newcommand{\eea}{\end{eqnarray}}
	\newcommand{\bit}{\begin{itemize}}
		\newcommand{\eit}{\end{itemize}}
	\newcommand{\rf}[1]{(\ref{#1})}
	
	\title{Features of Spacetime-Symmetry Breaking and \\the Standard-Model Extension
		in Riemann--Cartan Geometry}
	
	\author{R.\ Bluhm}
	
	\address{Department of Physics and Astronomy, Colby College,\\
		Waterville, ME 04901, USA}
	
	\begin{abstract}
		For over two decades, the gravity sector of the Standard-Model Extension (SME)
		has served as a phenomenological framework for testing
		spacetime symmetry breaking in the presence of gravity.
		During this time, various theoretical features have been examined in greater detail and some
		refinements have been made.  
		In particular, differences between spontaneous and explicit breaking of diffeomorphisms,
		local translations, and local Lorentz transformations in Riemann-Cartan geometry, 
		as well as their corresponding consistency issues with geometric and mathematical identities,
		have been probed more deeply.
		This has led to a modified version of the SME being developed 
		that is suitable for investigating explicit breaking in gravity theories,
		which can be used as well to search for new geometries that go beyond Riemann-Cartan.  
		A selective overview of some of these features is presented here.  
	\end{abstract}
	
	\bodymatter
	
	\section{Introduction}
	
	The gravity sector of the Standard-Model Extension (SME) contains all observer-invariant
	terms that couple gravitational and matter fields with fixed background fields associated
	with spacetime symmetry breaking.\cite{akgrav04}
	It is naturally defined using a vierbein formalism to accommodate fermion fields with spin,
	where the independent gravitational fields are the vierbein $\vb \mu a$ and spin connection $\nsc \mu a b$.  
	The spacetime can then have both curvature and torsion,
	and the geometry is Riemann--Cartan.
	The generic form of the action with an Einstein--Hilbert term can be written as
	\beq
	S =  \fr 1 {2 \ka} \int d^4x \, e \, R (\vb \mu a,\nsc \mu a b) 
	+  \int d^4x \, e \, {\cal L}_{g, m, \bar k_X}  (\vb \mu a,\nsc \mu a b, f^\ps, \bar k_X) \, ,
	\label{kECS}
	\eeq
	with $\ka = 8 \pi G$, and where $f^\ps$ represents matter fields with indices represented by $\ps$,
	while $\bar k_X$ is a fixed background with indices represented by $X$.
	In each of these, the indices can be combinations of spacetime indices (denoted as $\mu$, $\nu$, \ldots)
	and local indices ($a$, $b$, \ldots).
	
	The Einstein, torsion, and matter field equations of motion are obtained, respectively, by varying the action
	with respect to $\vb \mu a$, $\nsc \mu a b$, and $f^\ps$, and are given as
	\bea
	G^{\mu\nu} 
	= \ka \bar T_e^{\mu\nu} \, ,  \quad\quad
	\label{EinsteinEqkbar} \\
	\hat T^{\la\mu\nu}  
	=   - \ka \bar S_{\om \, \pt{\mu} }^{\pt{\om \, } \la\mu\nu} \, ,
	\label{TorsionEqkbar} \\
	\fr {\de S_{g, m, \bar k_X}} {\de f^\ps} = 0 \, ,
	\quad\quad\quad\,\,\,\,
	\label{feqkbar}
	\eea
	where
	\beq
	\hat T^{\la\mu\nu} = T^{\la\mu\nu}  
	+ T^{\al \pt{\al} \mu}_{\pt{\al} \al} g^{\la\nu}
	- T^{\al \pt{\al} \nu}_{\pt{\al} \al} g^{\la\mu}
	\eeq
	is the trace-corrected torsion tensor,
	$\bar T_e^{\mu\nu}$ is the energy-momentum,
	and $\bar S_{\om \, \pt{\mu} }^{\pt{\om \, } \la\mu\nu}$ is the spin density.
	The bars over the two latter quantities denote that they depend on the fixed background $\bar k_X$.
	
	As fixed backgrounds, the fields $\bar k_X$ do not transform under diffeomorphisms or particle Lorentz transformations,
	and as a result these symmetries are broken either explicitly or spontaneously.  
	In the case of explicit breaking, the backgrounds $\bar k_X$ are understood as completely nondynamical objects
	that are put directly into the action,
	while in the case of spontaneous breaking the backgrounds $\bar k_X$ arise as vacuum values of
	fully dynamical fields $K_X$, e.g., $\bar k_X = \vev{K_X}$.
	
	To maintain observer independence, the action $S$ must be invariant under general coordinate transformations
	and observer local Lorentz transformations.
	These invariances give rise to mathematical Noether identities that hold off shell.
	In addition, geometric Bianchi identities involving the curvature and torsion must always hold.
	Notice that the backgrounds $\bar k_X$ are not varied in the equations of motion shown above;
	however, they do transform under general coordinate transformations and observer Lorentz transformations.
	
	With the absence of variations of $\bar k_X$ in the equations of motion, 
	conflicts can arise between the off-shell Noether or Bianchi identities and
	the on-shell equations of motion for the vierbein, spin connection, and matter fields.
	This is highly problematic for the case of explicit breaking, because the variations with respect to $\bar k_X$
	need not vanish, so that
	\beq
	\fr {\de S_{\bar k_X}} {\de \bar k_X} \ne 0 \, .
	\label{nogo}
	\eeq
	In this case, theoretical inconsistency, referred to as no-go results, with Riemann--Cartan geometry can occur.\cite{akgrav04}
	In contrast, with spontaneous breaking,
	the variations of the action with respect to the dynamical fields $K_X$ for vacuum solutions $\vev{K_X} = \bar k_X$ do vanish,
	\beq
	\fr {\de S_{\bar k_X}} {\de \bar k_X} = 0 \, ,
	\label{SSB}
	\eeq
	since these are vacuum solutions,
	where $\vb \mu a$ and $\nsc \mu a b$ also equal their vacuum values
	$\vev{\vb \mu a}$ and $\vev{\nsc \mu a b}$.
	Thus, the no-go results are evaded for the case of spontaneous breaking.
	It is for this reason that the backgrounds in the original SME gravity sector,
	referred to as SME coefficients, are considered as arising from
	spontaneous breaking,
	whereas explicit breaking that is inconsistent in Riemann--Cartan geometry
	can be viewed as an indication of beyond-Riemann--Cartan geometry, 
	such as Finsler geometry.\cite{akgrav04}
	
	Many experimental tests of spontaneous Lorentz and diffeomorphism violation 
	have been made using the SME,\cite{akqbjt,datatables}
	and theoretical questions concerning, for example, the fate of the Nambu--Goldstone (NG) modes 
	were considered early on.\cite{rbak}
	Since then, many other theoretical features of both explicit and spontaneous breaking have 
	been explored, which have implications for the gravity sector of the SME.
	A very selective overview of some of these is given in the following sections.
	
	\section{Explicit breaking revisited}
	
	Interest in theoretical models with explicit spacetime symmetry breaking,
	such as in Ho\v{r}ava gravity or massive gravity,\cite{HorMass}
	prompted a re-evaluation of the no-go results for explicit breaking.\cite{rb1}
	In a nutshell, it was found that in some cases the no-go results can be evaded.
	This is because even though the variations in Eq.~\rf{nogo} do not hold as
	equations of motion for fixed nondynamical backgrounds, 
	the left-hand side of Eq.~\rf{nogo} can nonetheless vanish for other reasons.
	One is that with explicit breaking, there are more degrees of freedom,
	due to the loss of local gauge symmetries.
	Thus, there are additional vierbein and/or spin connection modes compared to
	the case when the spacetime symmetries are not broken.
	Alternatively, the vanishing of the expression in Eq.~\rf{nogo} can hold
	if certain constraints are imposed on geometrical quantities, 
	such as the curvature.\cite{rb1,rjsp,obn21}
	
	In general, the counting of degrees of freedom shows that the no-go results
	can be evaded at least in principle at nonlinear order.
	However, in some cases, as with an ansatz solution for the metric,
	the extra modes do not appear.
	Similarly, working in perturbation theory can cause the extra modes to
	be suppressed at linear order, leading to inconsistency in low-energy effective theories.
	Likewise, in theories where the no-go results impose constraints on the curvature,
	these conditions can often be so severe as to make a theory nonviable.  
	Thus, questions of consistency remain paramount in models with explicit breaking.
	
	With these considerations in mind,
	a new version of the SME for the case of explicit breaking has been developed.\cite{akzl}
	Its action generalizes the original SME action to include all observer-independent
	terms coupling gravitational and matter fields with fixed {\it nondynamical} backgrounds
	that explicitly break spacetime symmetries.   
	General arguments are made showing that at the level of perturbation theory
	the new terms in the action lead to inconsistency in Riemann--Cartan geometry.
	Thus, the interpretation with the new version of the SME is that it can be used
	as a phenomenological framework in gravity tests looking for explicit breaking,
	but any detection of a signal would likely indicate a new geometry that goes
	beyond Riemann--Cartan.
	
	An important caveat in using the new SME framework with explicit breaking is
	that couplings to different components of a background field must be treated as
	independent interactions.
	For example, for the case of a nondynamical background vector, 
	terms in the action coupling to spacetime components $\bar k_\mu$ and $\bar k^\mu$
	are physically distinct and must be investigated separately.
	This is because if both $\bar k_\mu$ and $\bar k^\mu$ are nondynamical
	they cannot be related by the physical metric $g^{\mu\nu}$.
	Similarly, local components $\bar k_a$ cannot be treated as being physically related to
	spacetime components $\bar k_\mu$, because the dynamical vierbein $\vb \mu a$
	cannot connect two nondynamical quantities.
	However, since the local metric $\et_{ab}$ is itself fixed,
	couplings to local components $\bar k^a$ and $\bar k_a$ can be treated as physically equivalent.
	Notice that this caveat can be ignored for the case of spontaneous breaking
	in the context of the original SME because there the backgrounds are vacuum values,
	which are physically linked by the vacuum values for the dynamical vierbein,
	e.g., $\bar k_\mu = \vev{\vb \mu a}\,\bar k_a$.
	
	As a result of the differences between nondynamical components, 
	e.g., $\bar k_\mu$, $\bar k^\mu$, and $\bar k_a$,
	the phenomenological framework resulting from the SME with explicit breaking
	has many more terms and coefficients that need to be tested compared to the SME with
	dynamical backgrounds resulting from spontaneous breaking.\cite{akzl}  
	Examples in cosmology with explicit breaking have also recently been considered.\cite{rss22,nan22}  
	
	\section{Local translations}
	
	Gravity theories in Riemann--Cartan geometry have three relevant local symmetries:
	diffeomorphisms, local translations, and local Lorentz transformations.
	However, only two of these are independent,
	since suitable combinations of diffeomorphisms and local Lorentz transformations can reproduce local translations, and vice versa.\cite{HehlRMP76,Blag02,bg13}
	Poincar\'e gauge theories of gravity work with translations and Lorentz transformations
	as the fundamental symmetries,
	while the gravity sector of the SME uses diffeomorphisms and Lorentz transformations
	as the independent pair of local symmetries.
	A natural question that then arises is whether the breakings of these local symmetries 
	are related as well.
	
	With spontaneous breaking of diffeomorphisms and local Lorentz invariance,
	it must be remembered that the symmetries are not actually broken, 
	but rather they become hidden with respect to vacuum solutions that do not
	obey the symmetries.
	In this case, when a tensor has a vacuum value with spacetime components 
	that break diffeomorphisms, the vacuum vierbein can be used to obtain the local
	components for the tensor, which spontaneously break local Lorentz transformations.\cite{rbak}
	This is true as well for nonconstant scalars,
	which have derivatives along spontaneously chosen directions in both the spacetime and local frames.
	
	Since local translations can be formed as combinations of diffeomorphisms and local Lorentz transformations,
	any tensor vacuum value that spontaneously breaks these symmetries
	will also spontaneously break local translations.
	Hence, any tensor background that arises from spontaneous breaking will break all three
	local spacetime symmetries.
	Note in particular that the vacuum value for the vierbein, $\vev{\vb \mu a}$, has both a spacetime
	and a local component,
	so it by itself spontaneously breaks all three local symmetries.
	
	In contrast, with explicit breaking, a nondynamical tensor is included directly in the action,
	which can break spacetime symmetries.
	In this case,
	the three local spacetime symmetries need not all break if one of them breaks.\cite{ccyb,rb2}
	This is largely due to the fact that the physical vierbein cannot relate spacetime
	and local components of a fixed background.
	For example, for a fixed vector, theories with couplings to components $\bar k_\mu$, $\bar k^\mu$, and $\bar k_a$
	must be treated as independent theories, as described previously,
	and thus can cause explicit breaking of different symmetries.  
	For example, constant components $\bar k_\mu$ explicitly break diffeomorphisms and local translations,
	but not local Lorentz transformations,
	while constant components $\bar k_a$ explicitly break local translations and local Lorentz transformations, 
	but not diffeomorphisms. 
	
	\section{Nondynamical scalar backgrounds}
	
	Explicit breaking with fixed nondynamical scalar background fields gives rise to some noteworthy features.\cite{rb1,rjsp}
	For example, consider a gravity theory with a set of fixed nondynamical scalars, $\bar \ph^A (x)$, with $A = 1,2,\dots, N$.
	The presence of such scalars in the action explicitly breaks diffeomorphism invariance.
	However, to be physically viable the theory must still be observer independent,
	which means the action must be unchanged under general coordinate transformations, $x^\mu \rightarrow x^\mu + \xi^\mu$,
	where $\xi^\mu$ are arbitrary infinitesimal spacetime vectors.
	Mathematical identities following from this
	are at odds with the explicit diffeomorphism breaking and Bianchi identities,
	leading to no-go results, unless the following condition holds:\cite{rb1}
	\beq
	\left[ -D_\mu \fr {\partial {\cal L}} {\partial (\partial_\mu \bar \ph^A)} + \fr {\partial {\cal L}} {\partial \bar \ph^A} \right] \partial_\nu \bar \ph^A = 0 \, .
	\label{cond1}
	\eeq
	Assuming that the scalars are not constant, so that  $\partial_\nu \bar \ph^A \ne 0$, 
	and assuming their derivatives are linearly independent as well,
	it follows that a stricter set of conditions must hold:
	\beq 
	-D_\mu \fr {\partial {\cal L}} {\partial (\partial_\mu \bar \ph^A)} + \fr {\partial {\cal L}} {\partial \bar \ph^A}  = 0 \, .
	\label{ELcond}
	\eeq
	Note that these have the form of the Euler--Lagrange equations
	that would hold dynamically if the scalars $\bar \ph^A$ were instead dynamical.
	Evidently, to evade the no-go results, the Euler--Lagrange equations for the scalars
	must still hold despite the fact that the scalars are not dynamical.
	
	Equation~\rf{ELcond} cannot follow from dynamical field variations of $\bar \ph^A$, 
	since the scalars are fixed and nondynamical.
	Instead, Eq.~\rf{ELcond} must hold for one of two reasons.
	
	The first is that the covariant derivatives in Eq.~\rf{ELcond} can involve 
	additional degrees of freedom in the metric because of the loss of diffeomorphism invariance.  
	With breaking associated with four local vectors $\xi^\mu$,
	this would allow up to four additional modes,
	which matches the number of equations in Eq.~\rf{ELcond} if $N =4$.
	This result helps explain why the St\"uckelberg trick works in theories with
	explicit diffeomorphism breaking.\cite{ags03}
	In the St\"uckelberg approach, a fixed background tensor is replaced by derivatives of four
	fixed St\"uckelberg scalar fields, $\bar \ph^A$, with $A = 0,1,2,3$.
	The scalars are then promoted to dynamical fields
	with equations of motion given by their Euler--Lagrange equations,
	which happen to match the conditions for evading the no-go results in Eq.~\rf{ELcond}.
	Thus, as long as the St\"uckelberg scalars obey their equations of motion,
	then the no-go results are evaded.
	However, this is not guaranteed.
	A theory with St\"uckelberg fields can still be inconsistent if solutions
	to Eq.~\rf{ELcond} do not exist.
	
	A second way for the conditions in Eq.~\rf{ELcond} to hold is if
	a constraint is imposed on the spacetime geometry.
	An example of this is provided by Chern--Simons gravity in four dimensions,
	which includes a nondynamical scalar in the action.\cite{rjsp}
	The condition for evading the no-go result in this case,
	which is equivalent to the Euler--Lagrange equation for the scalar,
	is that the gravitational Pontryagin density must vanish,
	$\star R R = R^{\al}_{\pt{\al}\be\mu\nu} R_{\pt{\be}\al}^{\be\pt{\al}\mu\nu} = 0$.
	
	A nondynamical scalar can also be used to construct theories with a hybrid form 
	of spacetime symmetry breaking involving both explicit and spontaneous breaking.\cite{rb15b,rs25}
	This occurs when a fixed nondynamical scalar is included in a potential that causes
	spontaneous breaking.
	For example, in bumblebee models with a vector field $B_\mu$, 
	spontaneous breaking is induced when the potential has the form 
	$V(B^\mu B_\mu + b^2)$ with $b$ equal to a constant.
	A vacuum solution $\vev{B_\mu} = b_\mu$ occurs,
	with a timelike vacuum value 
	that can be written as $b_\mu = (b,0,0,0)$ 
	in an appropriate frame.
	In this case, NG modes are excited that stay in the minimum of the potential with $V^\prime = 0$,
	while a massive mode occurs when $V^\prime \ne 0$.
	By replacing the constant $b$ by a nondynamical scalar in the potential $V$,
	for example using a time-dependent scalar $b(t)$ in $V(B^\mu B_\mu + b(t)^2)$,
	explicit diffeomorphism breaking occurs.
	However, the potential continues to induce
	spontaneous breaking but in this case with a time-dependent vacuum solution,
	$b_\mu (t) = (b(t),0,0,0)$.
	The constraint that must hold to evade the no-go results is $V^\prime = 0$,
	which also follows from the Euler--Lagrange equation for $b(t)$.
	This constraint eliminates the massive mode,
	leaving only NG modes as excitations in $B_\mu$.
	
	\section{Spin connection and torsion backgrounds}
	
	The roles of the spin connection and torsion have also been explored in greater detail
	in gravity theories with spacetime symmetry breaking.\cite{ccyb,rb2}
	Examples are given here for both explicit breaking as well as spontaneous breaking.  
	
	In models with explicit breaking,
	it has been shown that torsion can occur even when the spin density
	due to the matter fields is zero.
	This is different from the case of Einstein--Cartan theory in
	the absence of spacetime symmetry breaking,
	where the torsion $T^{\la\mu\nu}$ is directly proportional 
	to the spin density $S_{\om \, \pt{\mu} }^{\pt{\om \, } \la\mu\nu}$.
	Thus, if the spin density vanishes then so does the torsion.
	However, with nondynamical background fields that cause explicit breaking,
	the torsion equation is modified, as shown in Eq.~\rf{TorsionEqkbar}.
	Here, the bars over the torsion and spin density indicate
	interactions of the gravitational and matter fields with a
	nondynamical background.  
	In models with pure-gravity couplings to the background,
	the torsion can be nonzero even in the absence of matter.\cite{ccyb}
	
	In theories with spontaneous breaking of spacetime symmetries,
	it is common to assume that while the vierbein has a nonzero
	vacuum value, $\vev{\vb \mu a} \ne 0$, the spin connection has a
	vanishing vacuum value, $\vev{\nsc \mu a b } = 0$.
	This is a simplifying assumption, 
	since the torsion is related to the spin connection by
	\beq
	T_{\la\mu\nu} = \vb \la a \big[ (\partial_\mu \lvb \nu a + \lsc \mu a b \, \vb \nu b) 
	- (\mu \leftrightarrow \nu ) \big] \, .
	\label{toreq}
	\eeq
	Thus, when the vierbein has a constant Minkowski 
	vacuum value, e.g., $\vev{\vb \mu a} = \de^a_\mu$,
	the torsion has a vanishing vacuum value whenever the spin connection 
	also has a vanishing vacuum value.
	However,
	this assumption need not always hold.
	Models with $\vev{\nsc \mu a b } \ne 0$ can also be considered,
	where in this case the torsion and spin density can both have nonzero vacuum values.
	
	For example, consider a bumblebee model with torsion,
	where the field strength is given by
	\beq
	B_{\mu\nu} = D_\mu B_\nu - D_\nu B_\mu
	= \partial_\mu B_\nu - \partial_\nu B_\mu - T^\la_{\pt{\la}\mu\nu} B_\la \, .
	\label{Bmunu}
	\eeq
	In a vacuum solution with a constant value of $\vev{B_\mu} = b_\mu$,
	the vacuum field strength becomes
	\beq
	\vev{B_{\mu\nu}} = - \vev{T^\la_{\pt{\la}\mu\nu}} \,\vev{\vb \la a}\, \bar b_a \, ,
	\label{vacBmunu}
	\eeq
	the torsion vacuum value is
	\beq
	\vev{T_{\la \mu \nu}} = \vev{\om_{\mu\la\nu}} - \vev{\om_{\nu\la\mu}} \, ,
	\label{Torvev}
	\eeq
	and therefore $\vev{\bar S_{\om \, \pt{\mu} }^{\pt{\om \, } \la\mu\nu}} \ne 0$.
	Thus, if $\vev{\nsc \mu a b } \ne 0$, nonzero vacuum values can occur for
	all of the quantities $\vev{B_{\mu\nu}}$, $\vev{T_{\la \mu \nu}}$, 
	and $\vev{\bar S_{\om \, \pt{\mu} }^{\pt{\om \, } \la\mu\nu}}$
	even in the absence of excitations in $B_\mu$.
	
	Vacuum values for the spin connection can also affect how the NG modes
	appear with spontaneous spacetime breaking.
	The NG modes originate as virtual broken symmetry transformations
	that stay in the potential minimum,
	which can consist of broken diffeomorphisms, local translations, 
	or local Lorentz transformations.
	However, with a bumblebee potential only three NG excitations arise that stay in
	the minimum with $V^\prime = 0$.
	The remaining degrees of freedom aside from the gravitational excitations
	and a massive mode are gauge fixed,
	and how this is carried out depends on whether $\vev{\nsc \mu a b }$ vanishes or not.  
	This is because $b_\mu$ by itself breaks diffeomorphisms
	but not local translations when $\vev{\nsc \mu a b } = 0$;
	whereas with $\vev{\nsc \mu a b } \ne 0$,
	$b_\mu$ by itself breaks both diffeomorphisms and local translations.  
	Thus, with $\vev{\nsc \mu a b } = 0$, and after gauge fixing, 
	the NG modes are found to arise from $b_\mu$ as virtual local Lorentz transformations.
	In contrast, when $\vev{\nsc \mu a b } \ne 0$ and after gauge fixing,
	the NG modes in this case appear as a locked combination of both
	local translations and local Lorentz transformations.\cite{rb2}

\end{document}